\documentclass[symmetry,article,accept,pdftex,moreauthors]{Definitions/mdpi} 

\firstpage{1} 
\pubvolume{14}
\issuenum{8}
\articlenumber{1736}
\pubyear{2022}
\copyrightyear{2022}
\externaleditor{Academic Editor: {Sergei D. Odintsov}} 
\datereceived{27 July 2022} 
\dateaccepted{16 August 2022} 
\datepublished{19 August 2022} 
\hreflink{https://doi.org/10.3390/sym14081736} 

\usepackage{physics}

\Title{Effects of the Vertices on the Topological Bound States in a Quasicrystalline Topological Insulator}

\TitleCitation{Effects of the Vertices on the Topological Bound States in a Quasicrystalline Topological Insulator}

\Author{Simone Traverso 
 $^{1}$, Niccol\`o Traverso Ziani $^{1,2,}$* 
 Maura Sassetti $^{1,2,}$*}

\AuthorNames{Simone Traverso, Niccolò Traverso Ziani, Maura Sassetti}

\AuthorCitation{Traverso, S.; Traverso Ziani, N.; Sassetti, M.}

\address{%
$^{1}$ \quad Dipartimento di Fisica, Università Degli Studi di Genova, Via Dodecaneso 33, 16146 Genova, Italy; simone.traverso@edu.unige.it\\
$^{2}$ \quad CNR SPIN,
 Via Dodecaneso 33, 16146 Genova, Italy}

\corres{Correspondence: traversoziani@fisica.unige.it (N.T.Z.); sassetti@fisica.unige.it (M.S.)}

\abstract{The experimental realization of twisted bilayer graphene strongly pushed the inspection of bilayer systems. In this context, it was recently shown that a two layer Haldane model with a thirty degree rotation angle between the layers represents a higher order topological insulator, with zero-dimensional states isolated in energy and localized at the physical vertices of the nanostructure. We show, within a numerical tight binding approach, that the energy of the zero dimensional states strongly depends on the geometrical structure of the vertices. In the most extreme cases, once a specific band gap is considered, these bound states can even disappear just by changing the vertex structure.}

\keyword{higher order topological insulators; quasicrystals; bound states} 

\begin{document}

\section{Introduction}
\textls[-15]{The quest for quantum states with enhanced robustness to uncontrollable perturbations, such as noise, triggered the study of the topological phases of matter~~\cite{ti2}. From the point of view of electronics and spintronics~\cite{spin1,spin2,spin3,spin33,spin4}, the main focus is on non-superconducting systems, and in particular on topological insulators~\cite{PhysRevLett.95.226801,PhysRevLett.95.146802,ti1,ti3}. For superconducting spintronics~\cite{supspin1,supspin2,supspin3,supspin4,supspin5} and topologically protected quantum computation~\cite{comp1,comp2,comp3,comp4,comp5,comp6,comp7,comp8,comp9,comp10,comp11,comp12,comp13,comp14}, even through parafermions~\cite{para1,para2,para3,para4}, a prominent role is conversely played by topological superconductors~\cite{masatoshi17}.}

The physics is similar in the two cases: The systems have a gapped bulk, and bound states located at the physical boundaries with energy in such gaps. The bound states are predicted to enjoy, for topological reasons, enhanced robustness with respect to perturbations and are hence promising for quantum technological applications.

The robustness of such topological states is, however, debated and, in general, appears to be reduced with respect to the theoretical predictions. Prominent examples in this direction deal with the origin of the conductance fluctuations in quantum spin Hall systems, which is not perfectly quantized even when time-reversal symmetry is not explicitly broken~\cite{qsh1,qsh2,qsh3,qsh4,qsh5,qsh6,qsh7,qsh8,qsh9}, and with the difficulty in establishing the topological (Majorana) nature of the bound states in one-dimensional topological superconductors~\cite{md1,md2,md3,md4,md5,md6,md7,md8,md9,md10,md11,md12,md13,md14,md15,md16,md17,md18,md19,md20}.

To make progress, new paradigms for topological protection have hence recently been explored. A very promising one is {\em higher order topology}~\cite{h1,h2,h3,h4,h5}. 
{ A $d$-dimensional higher order topological insulator or superconductor is a system in which the topologically protected states live in $n$ dimensions less than the gapped bulk, with $1<n\leq d$. The index $n$ is referred to as the \emph{order} of the topological insulator.} Very recently, three-dimensional gapped systems with one-dimensional topological states have been realized experimentally~\cite{h6,h7}. On the other hand, proposals for the highly relevant case of two-dimensional bulk and zero-dimensional topological bound states have been put forward, but not experimentally~accessed. 

Among such proposals, a noteworthy one has been envisioned by S. Spurrier and N. Cooper~\cite{spurrier}. The main ingredient is a quasicrystalline {two-dimensional} lattice, generated by two layers of graphene-like lattices with a 30 degree relative rotation, and possessing 12-fold rotational symmetry. Each layer consists of a Haldane model and the two layers have opposite Chern number. Moreover, the edges of the bilayer structure they consider are of the armchair and bearded type. The result is that the system, thanks to the presence of the rotational symmetry, hosts topological {zero-dimensional} bound states, which are degenerate, isolated in energy from the others, and localized at the vertices of the structure. { The system is hence a higher order topological insulator with $d=2$ and $n=2$}. A subsequent work~\cite{sub} showed that these properties also characterize samples with armchair-zigzag edges. However, in this case, the energy at which the gap opens drastically changes. While in the armchair-bearded case the gap opens at the charge neutrality point, in the armchair-zigzag case it is substantially moved away from it. This observation poses questions about the sensitivity to details of the model.

In this context, the aim of the present work is to address some questions related to the robustness of the properties of the topological bound states with respect to lattice perturbations at the corners of the structure: Do these corner modes always exist in a given gap? Which is their energy? How strong is the protection granted to their degeneracy?

We will show how the energy, degeneracy and even existence of the predicted corner modes astonishingly depend upon the precise disposition of the lattice sites around the sample vertices. These results seem to narrow the path for a hypothetical experimental realization of the system proposed, which would require site-wise precision in the realization of the sample in order to achieve the HOTI phase close to the Fermi level.

The paper is organized as follows: In Section \ref{sec2} we introduce the model, both for how it concerns the lattice and for how it concerns the Hamiltonian. In Section \ref{sec3} we present and discuss our results, answering one by one the questions we addressed above. In Section \ref{sec4} we draw our conclusions.

\section{Model}\label{sec2}
In order to proceed with the discussion of the results achieved, we must first introduce the model of HOTI on a quasicrystal, which will be the object of our analysis.
The present section is organized in two parts: in the first one we introduce the system lattice, while in the latter we present the model Hamiltonian. In what follows, we will consider a square sample instead of a dodecagonal one, as is the case in \cite{spurrier}. Indeed, when dealing with the robustness of the bound states, the choice of a square sample provides major computational advantages, while leaving the physics qualitatively unchanged.

\subsection{Lattice}
\begin{figure}[H]
    \includegraphics[scale=1.45]{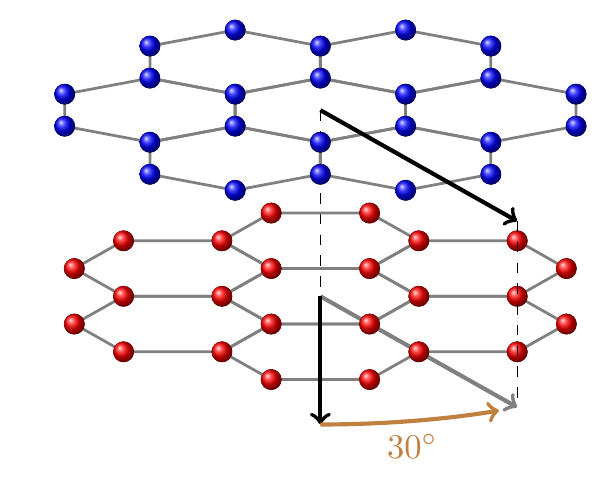}
    \caption{Scheme of the lattice construction: one starts from the AA stacking, and then rotates one of the two layers of $30^{\circ}$ around an axis passing through the centers of two superimposed hexagons.}
    \label{fig:lattice-scheme}
\end{figure}

Let us begin by describing the construction of a finite-size square sample of the quasicrystalline lattice under inspection: one starts with the two honeycomb layers superimposed with AA stacking (i.e., each atom of the top layer is right on top of an atom of the bottom layer) and rotates one of the two layers by thirty degrees with respect to the other, taking a rotation axis that goes through the center of two superimposed hexagons (see Figure~\ref{fig:lattice-scheme}). The resulting lattice is crystallographically equivalent to that of the \emph{graphene quasicrystal}, which has been shown to possess local bulk $C_{12}$ rotational symmetry~\cite{graphenequasicrystal}.

\begin{figure}[H]
    \includegraphics[width=0.8\textwidth]{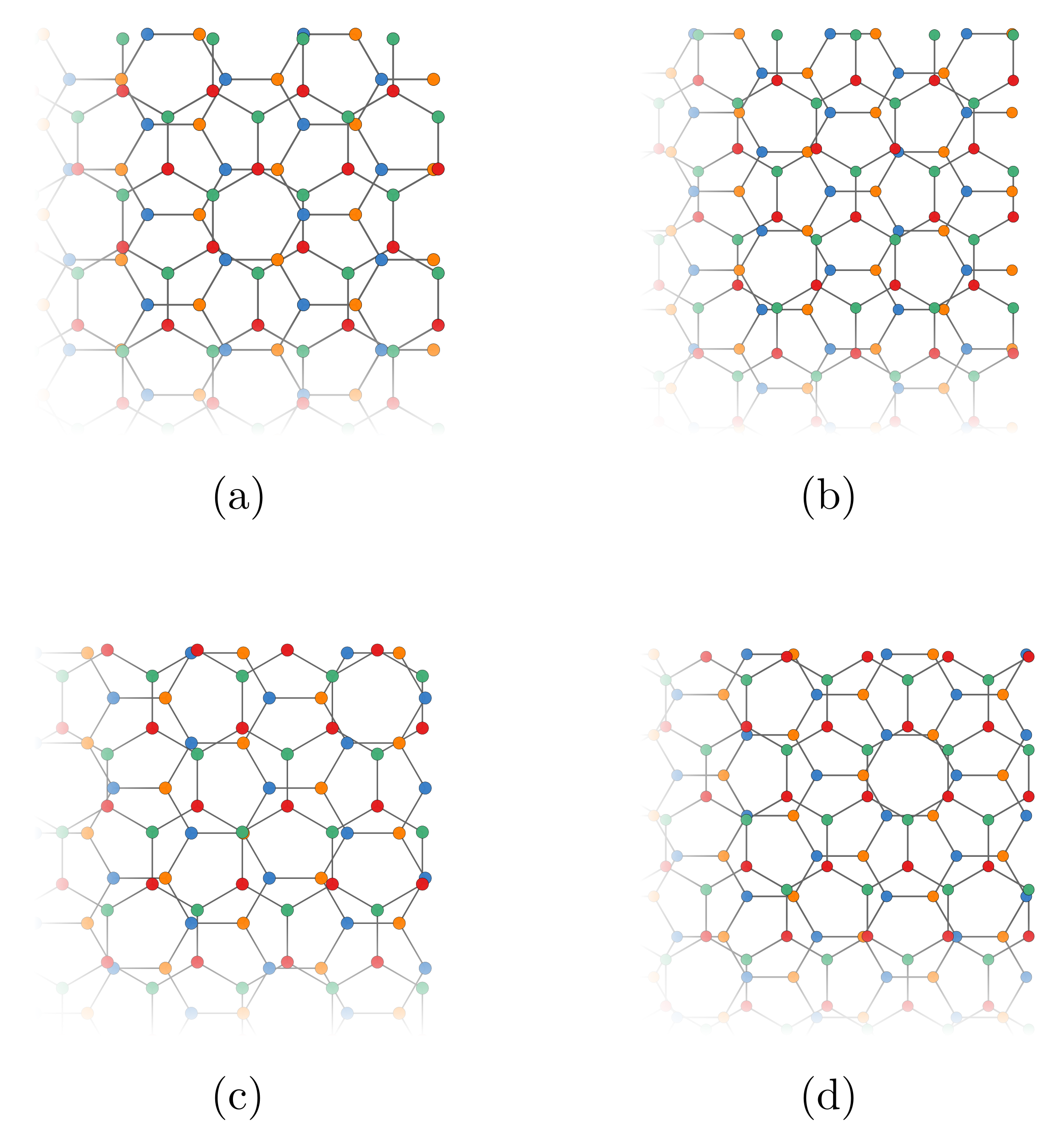}
    \caption{ A zoom around one vertex for each of the four possible lattice configurations discussed in the main text. Panels (\textbf{a},\textbf{b}) correspond to lattices presenting bearded-armchair edges, while Panels~(\textbf{c},\textbf{d}) correspond to lattices with zigzag-armchair edges.}
    \label{fig:01}
\end{figure}

We then crop our lattice with a square shape centered in the rotation center, and oriented in such a way that the edges are not jagged. 
As already noted in \cite{sub}, the edges of the resulting system---which possesses $C_4$ global rotational symmetry---can only be given by two possible pairs of the upper and lower layer edges: bearded-armchair or zigzag-armchair. However, even once the combination of the edges has been chosen, there are still two families of square lattices that can be obtained: these differ with each other for the disposition of the lattice sites around the vertices.
The zoom of a vertex for each of the four possible configurations just discussed is shown in Figure~\ref{fig:01}.
\vspace{-6pt} 

In \cite{sub}, it was shown how the properties of the model for an HOTI on the (spinless) graphene quasicrystal first proposed in \cite{spurrier}, depend on the shape of the edges. Here, we want to discuss how, even when the edge combination is fixed, the features of the HOTI phase can still depend on the specific configuration of the sites near the vertices. In order to approach this target, let us first introduce the model Hamiltonian.

\subsection{\label{subsec:ham} Model Hamiltonian}
\textls[-15]{The HOTI model we analyze \cite{spurrier} is governed by the following tight-binding Hamiltonian}
\begin{equation}
	H= t \sum_{\langle ij\rangle}c_i^\dagger\tau_0c_j +\lambda_H\sum_{\langle\langle ij\rangle \rangle}i\nu_{ij}c_i^\dagger \tau_z c_j +\lambda_{\perp}\sum_{ij}t^\perp_{ij}c_i^\dagger\tau_x c_j,
	\label{eq:spurrier_ham}
\end{equation}
where we denote $c_i=(c_i^{t},c_i^{b})^T$, with the spinless Fermionic operators $c_i^{\dagger,t/b}$ creating a fermion on the $i$-th lattice site of the top/bottom layer of honeycomb lattice. The Pauli matrices $\tau_i$ act in the space of layer pseudospin, coupling the Fermionic operators of the two layers. The Hamiltonian $H$ presents two in-plane hopping terms: the first one describes nearest neighbors ($\langle ij\rangle$) hoppings with amplitude $t$, while the second one involves next-nearest neighbors ($\langle\langle ij\rangle \rangle$) hoppings of amplitude $\lambda_H$. The latter is associated to a complex phase $i\nu_{ij}$ ($\nu_{ij}=\pm 1$), given by the usual Haldane coupling~\cite{haldane, bernevig2013topological}. Also, an inter-layer hopping term is present, parameterized by $t^\perp_{ij}$. Its explicit form is borrowed from tight-binding models for real twisted bilayer graphene (TBG) samples~\cite{PhysRevB.85.195458, PhysRevB.87.205404}.
\begin{equation}
    t_{ij}^\perp = t^\perp \exp \left(-\frac{|\vec{d}_{ij}|-d_{0}}{\delta}\right),
    \label{eq:t_perp}
\end{equation}
with $\vec{d}_{ij}$ being the vector that goes from the $i$ site on one layer to the $j$ site on the other one. We take the the parameters appearing in Equation~(\ref{eq:t_perp}) as in tight-binding models for real TBG samples as well~\cite{PhysRevB.85.195458, PhysRevB.87.205404}: $\delta = 0.184 a$ is the decay length of the transfer integral, $d_0=1.362 a$ the distance between the two layers and $t^\perp=-0.178 t$ a coupling coefficient.

All lengths are expressed in units of the in-plane next-nearest neighbor distance $a$. Finally, with the aim of tuning the interlayer coupling strength, we add the parameter $\lambda_\perp$ in Equation~(\ref{eq:spurrier_ham}) (the parameter $t_\perp$ is taken in such a way that if $\lambda_\perp=1$, then the value of the coupling for interlayer vertical hoppings matches the value used in realistic tight binding models for TBG \cite{PhysRevB.85.195458, PhysRevB.87.205404}).

In order to achieve a slightly more intuitive perspective about the model, it is worth noting~\cite{spurrier} that without the interlayer hopping term ($\lambda_\perp=0$), the Hamiltonian actually consists of two Haldane models~\cite{haldane} with opposite Chern number {($\pm 1$ in the upper/lower layer respectively)} and with a relative $30^\circ$ twist. This is almost the same as the Kane--Mele model~\cite{PhysRevLett.95.226801, PhysRevLett.95.146802}, just with the spin mapped into layer pseudospin. Indeed, with $\lambda_\perp=0$ the low energy spectrum would be gapless, with the edges of the two planes hosting a pair of counter-propagating modes- just as in the Kane--Mele model. What the interlayer hopping does is to couple together these two Haldane models: more specifically, as was shown in \cite{spurrier}, it gaps out the counter-propagating edge modes on the two layers, making it possible for the \emph{second order topological insulator} (SOTI) phase to occur. Such a phase is characterized by the presence of bound states localized at the corners of the system (\emph{corner modes}), degenerate among themselves and with energy laying inside the edge-gap.

The phenomenology just discussed is encoded in the effective low energy theory of the model derived in \cite{spurrier}. This theory is only based on the symmetries of the model and it leads to a few possible values for the energy of the corner modes inside the energy gap. For the present case of square sample ($C_4$ global symmetry), these are given by:
\begin{equation}
    \varepsilon_{\text{cm}}^{\text{th}}=\cos\left(\pi \frac{p}{4}\right), \qquad p \in \mathbb Z.
    \label{eq:low_en_pred}
\end{equation}

Having introduced all the necessary ingredients, we can now proceed to presenting and discussing our results.

\section{Results}\label{sec3}
We will now show the results obtained through the numerical diagonalization of Hamiltonian~(\ref{eq:spurrier_ham}) on different square samples of the quasicrystalline lattice. For the construction and diagonalization of the model, the \emph{python} package \emph{pybinding}~\cite{pybinding} has been used. The fixed parameters of the Hamiltonian were set as described above. We set $\lambda_\perp=2$, resulting in interlayer hoppings of double strength with respect to real TBG samples. This choice does not change the physics on a qualitative level, but it increases the gap width, reducing the hybridization of the corner modes for the same edge lengths. Also, we set $\lambda_H=0.3t$, so that the Haldane bulk gap is maximized~\cite{haldane}.

The present section is organized with the aim of separately answering the questions addressed in the introduction about the stability of the corner modes with respect to local geometrical perturbations of the lattice. We start by addressing the issue of the existence of the corner modes in a certain gap.

\subsection{Existence}
\begin{figure}[H]
    \includegraphics[width=\textwidth]{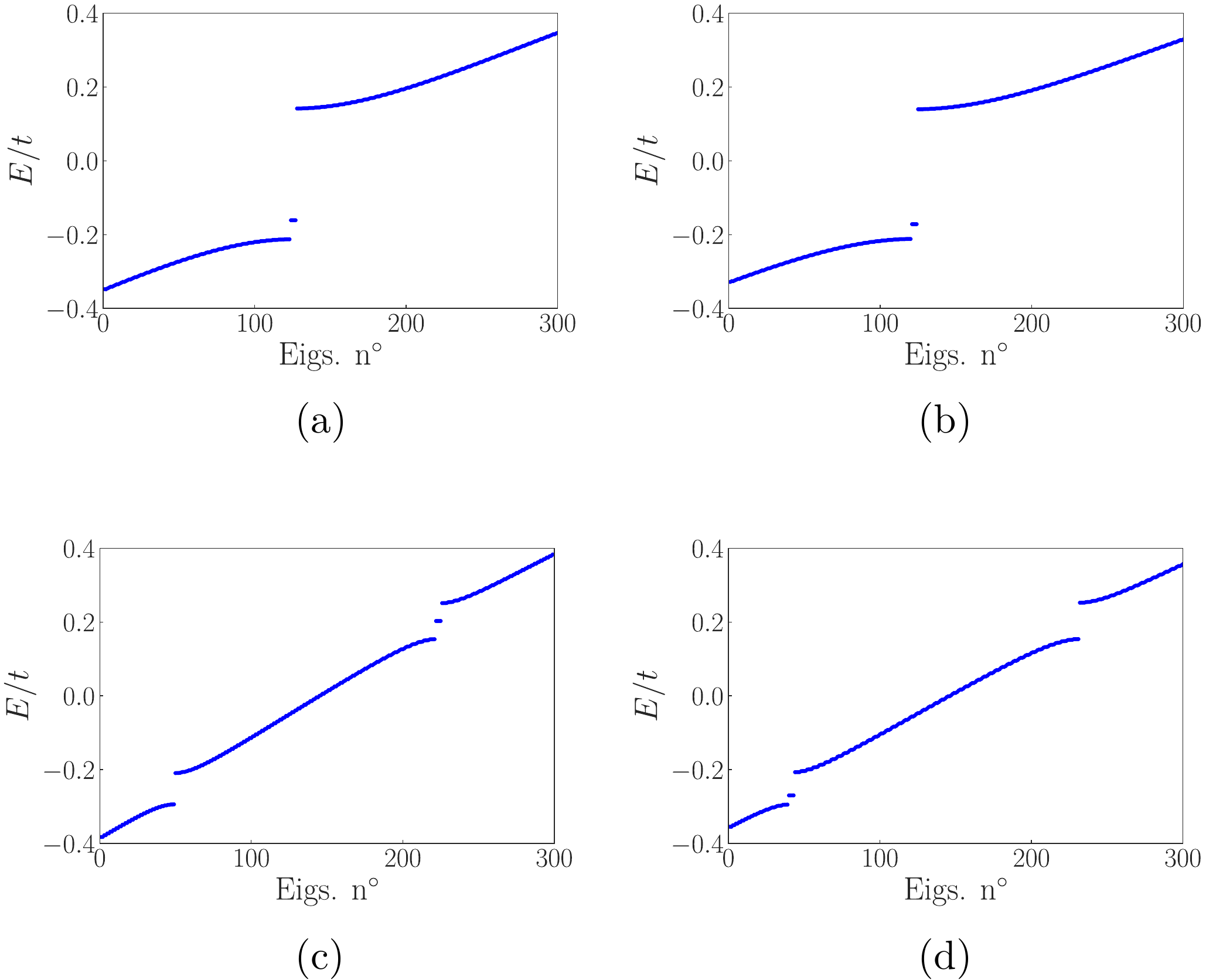}
    \caption{Eigenvalues obtained by diagonalizing the Hamiltonian in Equation~(\ref{eq:spurrier_ham}) for square lattices of four different sizes, one for each edge/vertex configuration in Figure~\ref{fig:01}. We set $\lambda_\perp=2$ and $\lambda_H=0.3t$, while all the other fixed parameters are as specified in the main text. The eigenvalues in Panel (\textbf{a}--\textbf{d}) correspond to samples with an edge of approximately $321 a$, $347 a$, $280 a$ and $306 a $ respectively.}
    \label{fig:02}
\end{figure}

Let us begin by presenting the low energy eigenvalues obtained for four different square samples, one for each of the categories in Figure~\ref{fig:01}. These are reported in Figure~\ref{fig:02}, where the plots of the eigenvalues are ordered the same way as the corresponding lattices, exemplified in Figure~\ref{fig:01}.

 Figure~\ref{fig:02}a,b reports the low-energy eigenvalues for two square samples with bearded-armchair edges and different shapes of the corners (see Figure~\ref{fig:01}a,b);
 in both cases we find four degenerate eigenvalues located inside the edge-gap. The existence of the topological bound states in the gap is hence robust with respect to variations in the corner shape. The lattices considered for the diagonalization present an edge with a length of $L\approx 321 a$ and $L\approx 347 a$, respectively. The slight difference in the energy of the degenerate quadruplet located inside the edge-gap will be discussed later in the article.

\textls[-25]{Figure~\ref{fig:02}c,d reports the low-energy eigenvalues for two square samples with zigzag-armchair edges} (in this case, the gap does not open at the charge neutrality point~\cite{sub}). In this case, the lattices considered for the diagonalization present an edge of length of $L\approx 280 a$ and $L\approx 306 a$ respectively. Just as before, these two lattices mainly differ for the disposition of the sites near the vertices, as can be understood by comparing Figure~\ref{fig:01}c,d. 
The results obtained by considering a lattice such as the one in Figure~\ref{fig:01}c or the one in Figure~\ref{fig:01}d are quite different. In the first case, as can be seen in Figure~\ref{fig:02}c, the quadruplet of degenerate eigenvalues is located right at the middle of the higher energy gap; on the other hand, in the latter case (Figure~\ref{fig:02}d) the quadruplet is located close to the bottom of the lower energy gap. This result demonstrates that, given a gap, the existence of topological bound states crucially depends on the shape of the vertices.

\begin{figure}[H]
    \includegraphics[width=0.65\textwidth]{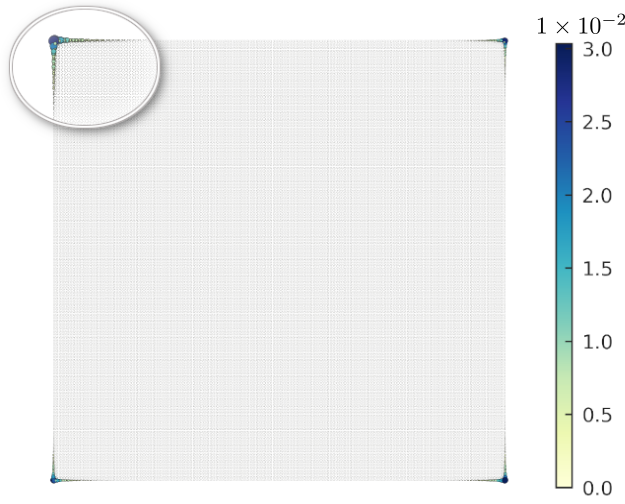}
    \caption{Probability density associated to the first in-gap eigenvalue of Figure~\ref{fig:01}a. On the top-left corner a zoom of the localization pattern of the bound state is reported.
    }
    \label{fig:bs}
\end{figure}

{Before proceeding, it is worth discussing the probability density related to the bound states. This is pictured in Figure~\ref{fig:bs}, for the first in-gap state in the set up of Figure~\ref{fig:01}a. All other eigenstates behave in a qualitatively analogous way. As expected, the bound states are located at the vertices of the structure. Moreover, their localization length is related to the difference in energy separating them to the lowest closest continuum energy states.}

From the analysis just reported, we learn that the existence of the eigenvalues associated to the corner modes in a certain gap is strongly dependent on the disposition of the lattice sites around the vertices in the armchair-zigzag case, while it is not in the bearded-armchair one.

\subsection{Energy}
By looking at the spectra in Figure~\ref{fig:02}, it is evident how the energy of the corner modes is strongly influenced by the disposition of the sites at the vertices of the system. In order to be a little more quantitative, we can try to compare the actual results reported above with the predictions of the low energy theory developed in Ref.~\cite{spurrier}. If we call $m$ the half-width of the gap and we refer with $E_{\text{mg}}$ to the midgap energy and with $E_{\text{cm}}$ to the energy of the corner modes, we can compute the normalized energy of each quadruplet with respect to the corresponding midgap as:
\begin{equation}
    \varepsilon_{\text{cm}}= \dfrac{E_{\text{mg}}-E_{\text{cm}}}{m}.
    \label{eq:en_norm}
\end{equation}

The values of such normalized energy for the four spectra in Figure~\ref{fig:02} are reported in Table~\ref{tab:01}.

\begin{table}[H]
\caption{Values of the normalized energy $\varepsilon_{\text{cm}}$, defined in Equation~(\ref{eq:en_norm}), for the spectra in Figure~\ref{fig:02}.}
 
   \setlength{\cellWidtha}{\columnwidth/3-2\tabcolsep-00in}
\setlength{\cellWidthb}{\columnwidth/3-2\tabcolsep-00in}
\setlength{\cellWidthc}{\columnwidth/3-2\tabcolsep-00in}
\scalebox{1}[1]{\begin{tabularx}{\columnwidth}{>{\PreserveBackslash\centering}m{\cellWidtha}>{\PreserveBackslash\centering}m{\cellWidthb}>{\PreserveBackslash\centering}m{\cellWidthc}}
\toprule
         & \boldmath{$L$} & \boldmath{$\varepsilon_{\text{cm}}$} \\
         \midrule
        (a) & $321 a$ & $-$0.708 \\
        (b) & $347 a$ & $-$0.772 \\
        (c) & $280 a$ & +0.010 \\
        (d) & $306 a$ & $-$0.430 \\
        \bottomrule
    \end{tabularx}}    
    \label{tab:01}
\end{table}

Recall that the prediction of the low energy theory for the energy of the corner modes was from Equation~(\ref{eq:low_en_pred}), $\varepsilon_{\text{cm}}^{\text{th}} = \cos\left(\frac{p\pi}{4}\right)$, with $p$ an integer. Comparing with the values in Table~\ref{tab:01}, we see that the cases (a) and (c) substantially meet the theoretical value for \mbox{$p=3$ \cite{spurrier}} and $p=2$ respectively. The other two configurations instead, lead to values of the normalized energy which are not compatible with the predictions of the low energy theory. With reference to Figure~\ref{fig:01}, we can observe that the cases (b) and (d) correspond to samples with lattice sites located just at the vertices. This leads us to the conclusion that the presence of lattice sites just at the corner may act as a sort of on site chemical potential term, causing a shift of the corner mode energies.

\begin{figure}[H]
    \includegraphics[width=\textwidth]{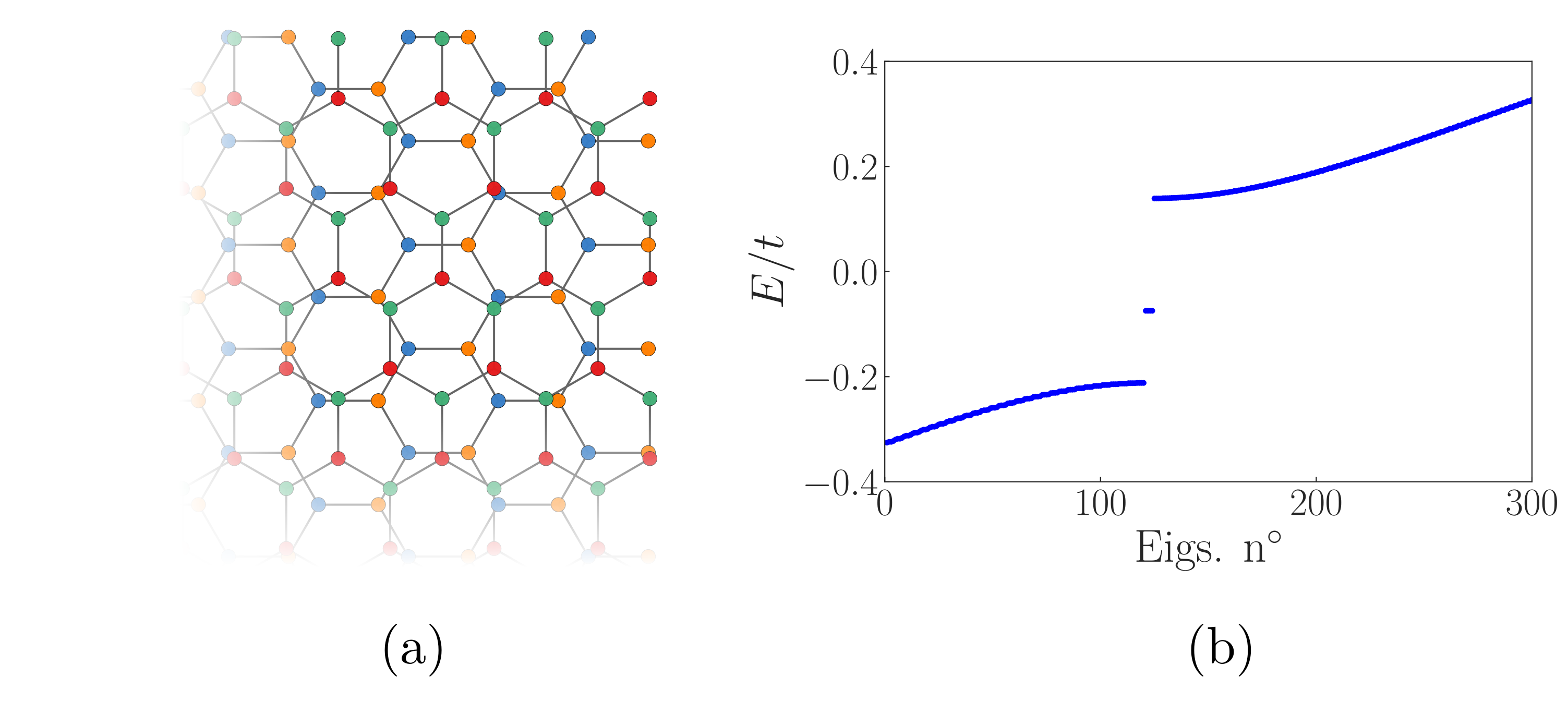}
    \caption{
    Panel (\textbf{a}): Zoom around the corner of a square lattice such as the one in Figure~\ref{fig:01}b, but with the corner sites removed. Panel (\textbf{b}): Eigenvalues obtained by diagonalizing the Hamiltonian in Equation~(\ref{eq:spurrier_ham}) for a square lattice with an edge of length $L\approx 347 a$, with the corner sites removed, as depicted in Panel (\textbf{a}). For the diagonalization we set $\lambda_\perp=2$ and $\lambda_H=0.3t$ as before and all other parameters as specified in the main text.}
    \label{fig:03}
\end{figure}

Indeed, this statement is corroborated by the results achieved considering the same lattice of Figure~\ref{fig:02}b, but with the lattice sites located right at the vertices removed, so that the resulting corners are as in Figure~\ref{fig:03}a. By diagonalizing the Hamiltonian in Equation~(\ref{eq:spurrier_ham}) for this specific lattice configuration, one finds the low energy spectrum reported in Figure~\ref{fig:03}b. Its comparison with the previous one in Figure~\ref{fig:02}b shows that the removal of the corner sites modifies the energy of the corner states, which is now visibly higher.

\subsection{Degeneracy}
\begin{figure}[H]
    \includegraphics[width=\textwidth]{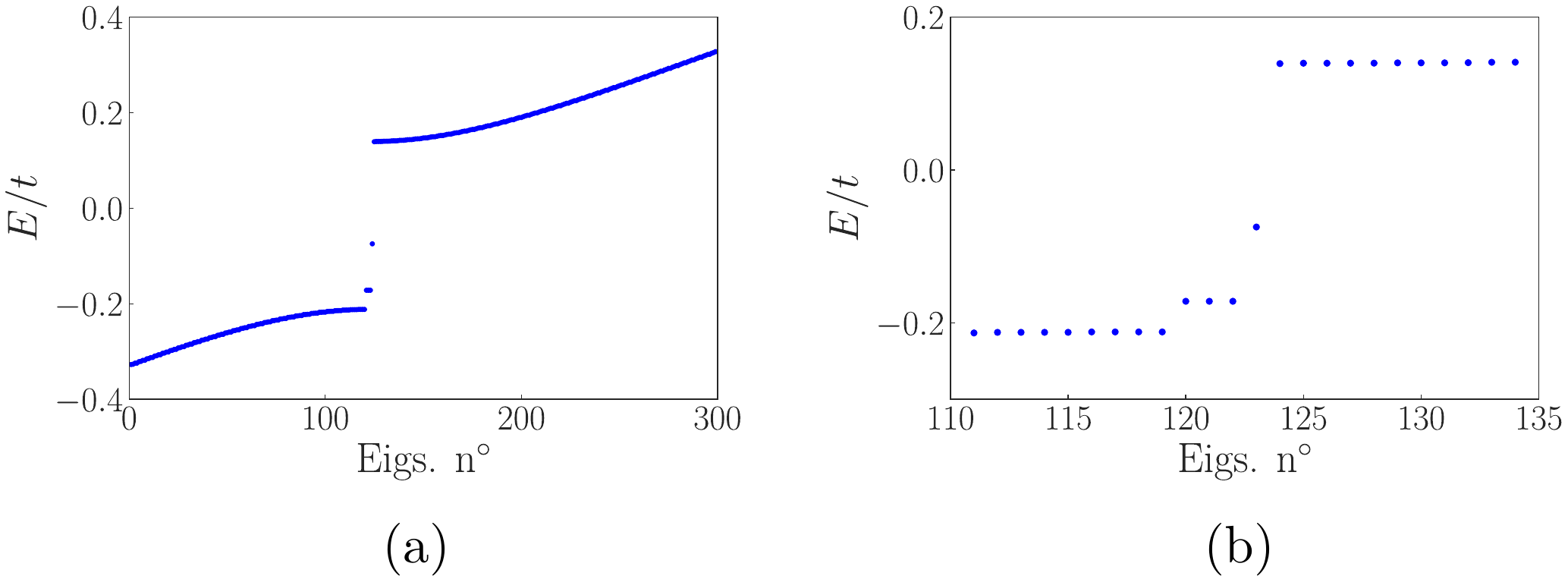}
    \caption{Panel (\textbf{a}): Eigenvalues obtained by diagonalizing the Hamiltonian in Equation~(\ref{eq:spurrier_ham}) for a square lattices with an edge of length $L\approx 347 a$, with the sites located at just one of the four corners removed, as in Figure~\ref{fig:03}a. For the diagonalization we set $\lambda_\perp=2$ and $\lambda_H=0.3t$ as before and all other parameters as specified in the main text. In Panel (\textbf{b}) a zoom of the spectrum around the gap.}
    \label{fig:04}
\end{figure}

In order to assess the degree of protection of the degeneracy of the bound states, we consider a square sample with bearded-armchair edges, with an edge of length $L\approx 347 a$, and we remove the lattice sites located at only one of its corners. In other words, the resulting lattice is just like the ones considered in order to obtain the spectra in Figures~\ref{fig:02}b~and~\ref{fig:03}b, though with three of the corners as in Figure~\ref{fig:01}b and with the remaining one as in Figure~\ref{fig:03}a. Strictly speaking, the system considered would now have lost the original exact $C_4$ symmetry, which is broken just because of the different disposition of the sites at one of the vertices. On the other hand, the $C_4$ symmetry is still present concerning the bulk and edges of the system. This should be enough to grant the validity of the low energy theory~\cite{spurrier}, at least for what it concerns the existence of the bound states. The spectrum resulting from the numerical diagonalization is shown in Figure~\ref{fig:04}.

One can clearly see that the original quadruplet of Figure~\ref{fig:02}b splits into a triplet plus a singlet. By carefully examining the precise energies of the eigenvalues, one finds that the triplet is at the energy of the quadruplet in Figure~\ref{fig:02}b (let us call it $\varepsilon$), while the singlet is at the energy of the quadruplet in Figure~\ref{fig:03}b (let us call it $\varepsilon'$). Moreover, one can also plot the probability density of the eigenstates associated with the in-gap eigenvalues (not shown here), finding that corner-modes corresponding to the three degenerate eigenvalues with energy $\varepsilon$ are localized on the three ``untouched'' vertices (Figure~\ref{fig:01}b), while the one with energy $\varepsilon'$ is localized on the cropped vertex (Figure~\ref{fig:03}a).

This analysis drives us to two main conclusions: first, the degeneracy of the corner modes is protected just as long as the $C_4$ symmetry is exact on the whole system (corners included). If we geometrically perturb the system at any vertex causing a local $C_4$ symmetry breaking, we destroy the degeneracy between the corner modes, though they still survive at different energies inside the gap- if originally present. Second, as we already noted in the previous subsections, the energy of the corner modes inside the edge gap crucially depends on the disposition of the lattice sites at the corners.

\section{Conclusions}\label{sec4}
Our article deals with the quasicrystalline higher order topological insulator predicted in Equation~(\ref{eq:spurrier_ham}). We address three questions related to the properties of the topologically protected zero-dimensional bound states characterizing the model. Such questions are related to the robustness with respect to changes in the structure of the vertices and are the following ones: Do the topological bound states always exist in a given gap? Is their energy fixed? Is their degeneracy protected? We show that the existence of the bound states in a given energy gap is independent of the structure of the vertices if the edges are of the armchair-bearded type, while it is not in the armchair-zigzag case. On the other hand, the energy of the bound states crucially depends on the structure of the vertices in all cases, and the degeneracy is consequently not protected.
We hence conclude that if the system is intended to be used as a higher order topological insulator, a precise control of the lattice sites at the vertices is needed.


\vspace{6pt} 

\authorcontributions{Conceptualization, N.T.Z. and M.S.; Programming and calculations S.T. All
authors contributed to the interpretation of the results and to the preparation of the draft. All authors
have read and agreed to the published version of the manuscript.}

\funding{This work was supported by the ``Dipartimento di Eccellenza MIUR 2018-2022''.}
\institutionalreview{Not applicable 
}

\informedconsent{Not applicable 
}

\dataavailability{Not applicable
}

\conflictsofinterest{The authors declare no conflict of interest.}






\begin{adjustwidth}{-\extralength}{0cm}
\reftitle{References}

\end{adjustwidth}
\end{document}